\documentclass[twocolumn,prb,aps,nobalancelastpage,showpacs,citeautoscript]{revtex4-1}

\usepackage{amsmath}
\usepackage{graphicx}
\usepackage{bm}

\begin{document}

\title{Location of gap nodes in the organic superconductors $\kappa$-(ET)$_2$Cu(NCS)$_2$ and
$\kappa$-(ET)$_2$Cu[N(CN)$_2$]Br determined by magnetocalorimetry}

\author{L. Malone,$^1$ O.J. Taylor,$^1$ J.A. Schlueter,$^2$ and A. Carrington$^1$}
\affiliation{$^1$ H. H. Wills Physics Laboratory, University of Bristol, Tyndall Avenue, Bristol, UK.\\$^2$ Materials
Science Division, Argonne National Laboratory, Argonne, Illinois 60439, USA.}

\date{\today}
\begin{abstract}
We report specific heat measurements of the organic superconductors $\kappa$-(ET)$_2$Cu(NCS)$_2$ and
$\kappa$-(ET)$_2$Cu[N(CN)$_2$]Br.  When the magnetic field is rotated in the highly conducting planes at low
temperature ($T\simeq 0.4$\,K), we observe clear oscillations of specific heat which have a strong fourfold component.
The observed strong field and temperature dependence of this fourfold component identifies it as originating from nodes
in the superconducting energy gap which point along the in-plane crystal axes ($d_{xy}$ symmetry).
\end{abstract}

\maketitle

\section{Introduction}

The organic charge transfer salts with general formula $\kappa$-(ET)$_2$X [where ET stands for
bis(ethylenedithio)-tetrathiafulvalene] have attracted considerable interest because of their unconventional properties
\cite{Lang03}.  Like the high-$T_c$ cuprates this family of quasi-two-dimensional materials exhibit a low temperature
superconducting ground state which is in close proximity to an antiferromagnetically ordered Mott insulating state. The
position of the various members of the series in the phase diagram is determined by the `chemical pressure' exerted by
the anion X. There is considerable evidence that the superconductivity is unconventional \cite{Lang03}. Power-law
temperature dependencies observed in thermal conductivity \cite{Belin_4728_1998}, NMR \cite{DESOTO_10364_1995},
magnetic penetration depth \cite{Carrington_4172_1999} and recently specific heat measurements
\cite{Taylor_057001_2007} point strongly to there being nodes in the superconducting energy gap in certain directions
of $\bm{k}$-space. In order to gain a better insight into the mechanism for the superconductivity it is important to
know the location of these nodes.

The two most widely studied organic superconductors are $\kappa$-(ET)$_2$Cu(NCS)$_2$  and
$\kappa$-(ET)$_2$Cu[N(CN)$_2$]Br (which we abbreviate to $\kappa$-NCS and $\kappa$-Br respectively) as these have the
highest superconducting transition temperatures (T$_c$ of $\sim$ 9.5\,K and  $\sim$12.5\,K respectively) at ambient
pressure. Angle dependent magneto-thermal conductivity \cite{Izawa_027002_2002} measurements indicate that in
$\kappa$-NCS the gap symmetry is d$_{x^2-y^2}$ (i.e., with nodes at 45$^{\circ}$ to crystal axis).  This conclusion is
also supported by angle dependent tunnelling measurements \cite{Arai_104518_2001}, however it is at odds with most
theories of superconductivity such as that of Schmalian \cite{Schmalian_4232_1998} based on spin-fluctuation mediated
pairing (for a review see Ref.\ \onlinecite{Kuroki06}) which predict a $d_{xy}$ pairing state (i.e., with nodes along
the crystal axes).  In this case the nodes are in the same location, with respect to the dominant nearest neighbor
antiferromagnetic coupling direction, as in the high $T_c$ cuprates.  However, Kuroki {\it et al.}
\cite{Kuroki_100516_2002,Kuroki06} calculate that a d$_{x^2-y^2}$  pairing state is often close in energy to the
$d_{xy}$ and can dominate for certain model parameter values.

Here we report an investigation of the location of the nodes in the order parameter of both $\kappa$-NCS and
$\kappa$-Br using angle dependent magnetocalorimetry as a probe. The magnetic field was rotated in the highly
conducting plane of the sample and oscillations of the specific heat were observed. The location of the maxima and
minima of the fourfold component of these oscillations points to the order parameter having $d_{xy}$ symmetry. This
appears to contradict the thermal conductivity results but recent sophisticated theories of magneto-oscillatory
specific heat and thermal conductivity predict rather complicated phase diagrams, with the phase of the oscillations
with respect to the nodes changing in distinct regions of field and temperature space.  Hence, the interpretation of
the experimental data is not straightforward and indeed these two seemingly opposite conclusions are not necessarily
incompatible.

Angle dependent magneto-specific heat oscillations were first used successfully to locate the minima in the
superconducting energy gap of YNi$_2$B$_2$C \cite{ParkSCKL03}. Subsequently, experiments were performed in several
heavy fermion supercondutors: Sr$_2$RuO$_4$ \cite{Deguchi_047002_2004}, CeColn$_5$ \cite{AokiSSSOMM04}, CeIrIn$_5$
\cite{Kasahara_207003_2008}, PrOs$_4$Sb$_{12}$ \cite{SakakibaraYCYTAM07} and URu$_2$Si$_2$ \cite{SakakibaraYCYTAM07} as
well as the strongly anisotropic $s$-wave superconductor CeRu$_2$ \cite{SakakibaraYCYTAM07}. In the simplest case, the
oscillations arise from the field induced `Doppler' shift of the energies of the quasiparticle states
\cite{Vekhter_R9023_1999}, $\delta E \propto \bm{v_s\cdot k}$, where $\bm{v_s}$ is the velocity of the screening
currents ($\propto H$) and $\bm{k}$ is the quasiparticle momentum. Close to a node or deep gap minimum, this field
induced shift can cause a substantial change in the population of the quasiparticle energy levels. In the case of a
simple two dimensional $d$-wave superconductor, if the field is applied along a node then only two nodes will
contribute to the change in the density of states as the field is perpendicular to the other two. However, if the field
is applied at 45$^{\circ}$ to this direction all four nodes will contribute, leading to an increase in the angle
averaged density of states. Hence, at sufficiently low temperature and field, the direction of the maxima in the
density of states should indicate the antinodal directions \cite{Vorontsov_237001_2006}. Although several experimental
quantities are sensitive to these oscillations, the specific heat is perhaps the most direct and simplest to interpret.
For example, to interpret the oscillations in the thermal conductivity, scattering of the quasiparticles from the
vortex lattice needs to be taken into account \cite{Vorontsov_237001_2006}.

\section{Experimental details}

Our samples were grown using a standard electrochemical method \cite{KiniGWCWKVTSJW90} and weighed $\sim$500\,$\mu$g.
The crystal orientation was determined by x-ray diffraction. Specific heat was measured using a purpose built
calorimeter\cite{Taylor_057001_2007} based on a bare chip Cernox \cite{lakeshore} thin film resistor.  The samples were
cooled slowly ($\sim 0.2$\,K/min) to avoid any stress induced phase separation \cite{Taylor_060503_2008}. The Cernox
was used as both heater and thermometer. Two different methods of measurement were used. To measure the temperature
dependence of $C$ the long-relaxation technique was used \cite{WangPJ01,Taylor_057001_2007}. For the rotation studies,
where much smaller temperature excursions from the base temperature are required, a thermal modulation technique was
used \cite{SullivanS68,Kraftmakher02}. An ac current ($\omega\simeq$ 3--6\,Hz) was passed through the Cernox and the
signals at $\omega$ and $3\omega$ detected with lock-in amplifiers. The $3\omega$ signal is inversely proportional to
the specific heat provided that $\omega$ is selected appropriately \cite{SullivanS68,Kraftmakher02}.  The thermometers
were calibrated in field using a $^3$He vapor pressure thermometer (below 4\,K) and a capacitance thermometer at higher
temperature. The angle dependence of the thermometer magnetoresistance was measured directly, and was found to be
negligibly small. Any angle dependence of the addenda specific heat was checked by measuring a pure Ag sample at our
base temperature and in fields up to 14\,T. No change was detected to within a precision of 0.2 \%. The thermal modulation
method has the advantage of high resolution at the expense of a small ($\sim$10\,\%) systematic error in the absolute
values.

For measurements as a function of field angle the calorimeter was mounted on a mechanical rotator whose rotation plane
was parallel to the field. Sample alignment was done by eye and was checked using optical images. Using this
arrangement we estimate that the field was kept parallel to planes to within a few ($\lesssim 5$) degrees. Experiments
were done either by rotating the sample at low temperature or by heating the sample above $T_c$ after each rotation. In
general, the two methods gave very similar results however there were occasional reproducible jumps in the data in
certain field directions when the sample was not heated above $T_c$ after each rotation, presumably because of vortex
pinning effects.

\begin{figure}
\includegraphics[width=7cm]{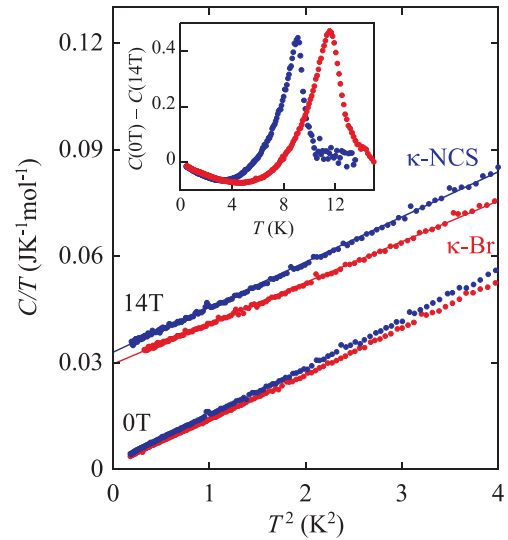}
 \caption{(Color online.) The temperature of the heat capacity in zero field and $B^\perp =14$\,T for samples of $\kappa$-Br and $\kappa$-NCS
  plotted versus $T^2$.  The inset shows zero field specific heat data (with normal state contribution at $B^\perp=14$\,T subtracted.)}
 \label{FigTdep}
\end{figure}

\section{Results and Discussion}

For $B^\perp$=14\,T (applied along the interlayer direction) both samples are in the normal state and the data
 follow $C/T = \gamma+\beta_3T^2+\beta_5T^4$, with $\gamma=29\pm$ 1\,mJK$^{-2}$mol$^{-1}$ for $\kappa$-Br
and $33\pm$ 1\,mJK$^{-2}$mol$^{-1}$ for $\kappa$-NCS in good agreement with previous measurements
\cite{Taylor_057001_2007}. We do not observe any Schottky-like upturns even at high field, so the $B^\perp$=14\,T data
is likely to be very close to the normal state $C$ at zero field \cite{Taylor_057001_2007}.  The zero field data with
the 14\,T data subtracted shown in Fig.\ \ref{FigTdep} shows the superconducting anomalies of each sample with
midpoints of 9.6\,K and 12.5\,K for $\kappa$-NCS and $\kappa$-Br respectively.

\begin{figure}
\includegraphics[width=8cm]{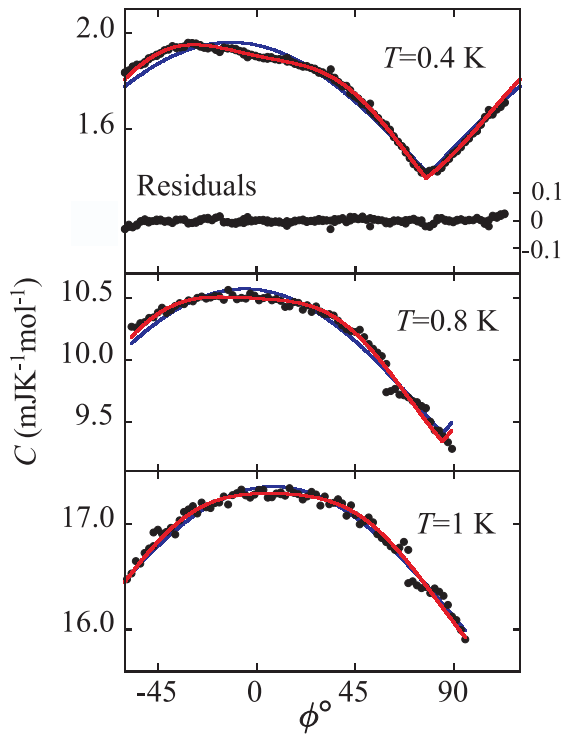}
 \caption{Color online. Raw data for $\kappa$-Br taken at three temperatures 0.4\,K, 0.8\,K and 1\,K.  The blue lines are a fit to
 a twofold angle dependence whereas the red lines include the additional fourfold term as described in the text.  }
 \label{FigRawAngleDep}
\end{figure}

\begin{figure}
\includegraphics[width=7cm]{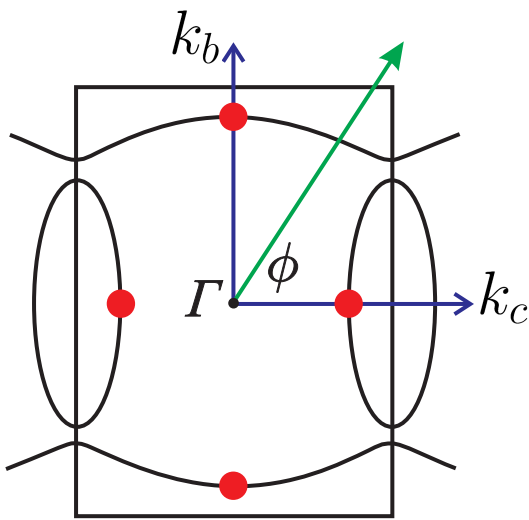}
 \caption{Color online. Sketch of the Fermi surface of $\kappa$-NCS \cite{Xu95}, along with the definition of $\phi$. The red dots show the position of the nodes in for
the $d_{xy}$ pairing state.}
 \label{fsfig}
\end{figure}

We now discuss the data with the field rotating in the basal plane. Fig.\ \ref{FigRawAngleDep} shows the raw $C(\phi)$
data for $\kappa$-Br at several temperatures and $B^\|$=3\,T which is $\sim$10\% of $H_{c2}$ for this field orientation
\cite{Shimojo2001427} ($\phi$ is the angle measured  relative to the in-plane $c$-axis).  Focusing on the lowest
temperature data $T$=0.4\,K, we see that the total change in $C$ with angle is around 33\% of the total (minus addenda)
or around 55\% of the electronic term at this field (the phonon contribution at $T=$0.4\,K $C_{\rm
Phonon}=0.85$\,mJ\,mol$^{-1}$K$^{-1}$ which is 60\% of the total). The largest component has twofold symmetry, however
to get a good fit we need also to include a fourfold term. We find the data are best fitted by the function
\begin{equation}
C(\phi)=C_0 + |C_2\cos(\phi+\delta_2)|+C_4\cos(4\phi+\delta_4) \label{cphifit}
\end{equation}
which fits the data perfectly within the noise as shown by the residuals displayed in the figure. The fit parameters
are $C_2=0.54$ mJ\,mol$^{-1}$K$^{-1}$, $\delta_2=13\pm 2^\circ$,  $C_4=0.04$ mJ\,mol$^{-1}$K$^{-1}$, and $\delta_4=0\pm
2^\circ$.

The $C_2$ two-fold term likely originates from a combination of sample misalignment and the anisotropy of the Fermi
surface of these compounds.  If the sample is misaligned by an angle $\varphi$ to the plane of rotation, then there
will be a component of magnetic field $B^\perp$ perpendicular to the planes given by $B^\perp/B_0 = \cos \phi \sin
\varphi$, where $B_0$ is the applied field and $\phi$ is the in-plane rotation angle.  As $H_{c2}$ is much lower for
field applied perpendicular rather than parallel to the planes there will be a induced component to $C(\phi)$ which
will depend on $B^\perp$. For a pure $d$-wave superconductor at low field we would expect that at low fields $\Delta
C(B)\sim (B/B_{c2})^{\frac{1}{2}}$, however in general the functional dependence may not be a square root (for example
if the field is less than a scale set by the impurity scattering bandwidth then $\Delta C \simeq
-B/B_{c2}\ln(B/B_{c2}$) Ref.\ \onlinecite{BarashSM97}). Experimentally, for these compounds for $B^\perp\gtrsim 0.1$\,T
$\Delta C(B)$ is close to linear \cite{Taylor_057001_2007}.  Given that this misalignment term must be even with
respect to $B$, the simplest form is then $\Delta C(\phi)= A \mathcal{F}|\cos(\phi + \delta)|$, where $A$ is constant
which depends on the misalignment angle and the form of the function $\mathcal{F}$ and $\delta$ is a constant which
depends on the misalignment angle. For simplicity we took $\mathcal{F}$ to be a simple power law so the misalignment
term is $\Delta C(\phi)= A |\cos(\phi + \delta)|^n$. A least squares fit to the data with this in place of the $C_2$
term in Eq.\ \ref{cphifit} gives $n=1.1\pm 0.1$, so we fixed $n=1$ for all temperatures and fields.

The crystal structures of $\kappa$-Br and $\kappa$-NCS are orthorhombic and monoclinic respectively. The Fermi surface
of $\kappa$-NCS has been determined by both tight-binding \cite{OSHIMA_938_1988} and first-principle \cite{Xu95}
calculations and quantum oscillation measurements \cite{OSHIMA_938_1988} and is sketched in Fig.\ \ref{fsfig}. The
quasi-two-dimensional Fermi surface would have a near ellipsoidal cross-section if all the ET-dimers were equivalent.
However, the difference in the inter-dimer hopping integrals causes a gap to appear where the ellipse crosses the
Brillouin zone boundary.  The Fermi surface of $\kappa$-Br is very similar to that of $\kappa$-NCS except that there
are twice the number of Fermi surface sheets because the unit cell contains twice the number of formula units due to
the doubling of the $c$-axis \cite{ChingXJL97}.

Theoretical calculations based on spin-fluctuation pairing suggest that the pairing state has either $d_{xy}$ or
$d_{x^2-y^2}$ symmetry.  For the case of $d_{xy}$ pairing as the gap has nodes along the crystal axes
\cite{Schmalian_4232_1998,Kuroki06} so the angle between these nodes remains $90^\circ$ even in the presence of the
orthorhombic distortion, and maxima in $C(\phi)$ with 4-fold symmetry as in Eq.\ \ref{cphifit} are expected.   In the
semiclassical theory the field dependence of the specific heat depends on the Fermi velocity and gradient of the
superconducting energy gap $\Delta$ near the nodes $d \Delta/d\phi|_{\rm node}$  \cite{Vekhter_R9023_1999}. So an
orthorhombic distortion of the Fermi surface can produce a two-fold $|\cos \phi|$ dependence of $C(B)$ similar to the
case of field misalignment. Indeed, in the field angle dependent thermal conductivity measurements of $\kappa$-NCS a
two fold term around two times larger than the four fold term was observed at $T=0.43$\,K (Ref.\
\onlinecite{Izawa_027002_2002}) even though the field was aligned to better than $0.01^\circ$ .

As the phase of the two-fold term in our measurements implies symmetry far from a crystal symmetry direction, this
suggests that a sizeable fraction of it originates from misalignment, however the thermal conductivity results suggest
that the intrinsic contribution is not negligible. In principle, in-situ adjustment of the crystal orientation (using
for example a vector magnet as in Ref.\ \onlinecite{Izawa_027002_2002}) and angle detection using the magnetoresistance
of the sample would allow us to distinguish between these two contributions however this was not possible with our
current set-up.

\begin{figure}
\includegraphics[width=8cm]{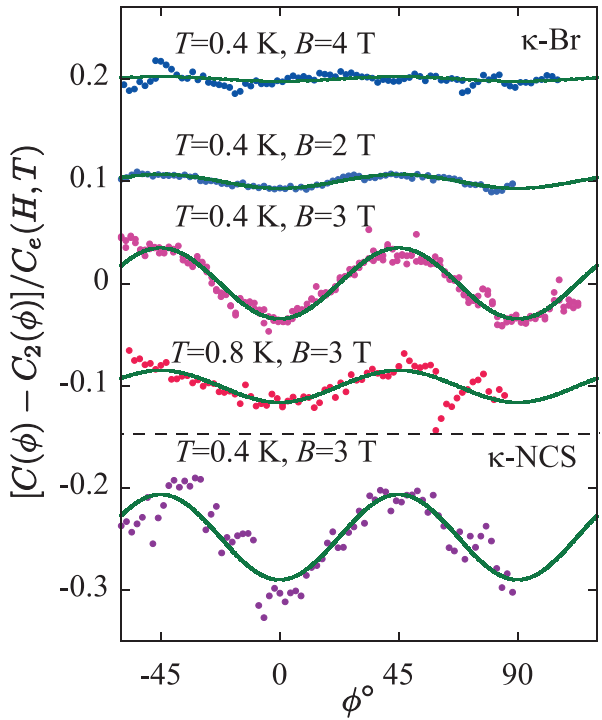}
 \caption{Color online.  Angle dependent heat capacity data with the twofold terms subtracted.
 The angle dependent changes $(C(\phi)-C_2(\phi))$ are divided by the angle averaged electronic specific heat
 at each temperature and field $C_e(H,T)$. The solid lines are fits to $C_4\cos(4\phi+\delta_4)$.}
 \label{Figfourfold}
\end{figure}

In Fig.\ \ref{Figfourfold} we show the  $\kappa$-Br data with the twofold term subtracted and normalized to the angle
averaged value of the electronic specific heat at the relevant temperature and field $C_e(H,T)$.  At $T$=0.4\,K the
fourfold term is now clearly evident. It has a relatively large peak to peak amplitude of $\sim$7\% of $C_e(H,T)$. Note
that the in-plane shape of this sample is an irregular polygon and so the fourfold term cannot result from shape
dependent vortex pinning effects.  The phase of this term is $\delta_4 = 0\pm$2$^\circ$, so that the minima occur when
the field is along the $a$ and $c$ crystal axes (equivalent to the $b$ and $c$ axes for $\kappa$-NCS in Fig.\
\ref{fsfig}). Hence, we conclude that the nodes in the gap function are located along the crystal axes, i.e., the
energy gap has $d_{xy}$ symmetry.

As the temperature is increased the relative size of the fourfold term compared to the angle average decreases rapidly,
and is barely discernable in the $T$=1\,K data. This strong decrease with increasing temperature points strongly to the
origin of the fourfold oscillation being from nodal quasiparticles, rather than, for example, $v_F$ anisotropy which
would have a much weaker $T$ dependence. For CeRu$_2$, which has a strongly anisotropic $s$-wave gap ($\Delta_{\rm
max}/\Delta_{\rm min} \simeq 5-10$), it is found that $C_4/C_e$ \textit{increases} with increasing $T$ up to a maximum
at $T/T_c\simeq 0.16$ \cite{SakakibaraYCYTAM07}. Fig.\ \ref{Figfourfold} also shows that the magnitude of the fourfold
term depends strongly on the applied field. It is maximal at $B$=3\,T and is significantly reduced for $B$=2\,T and
4\,T.

\begin{figure}
\includegraphics[width=8cm]{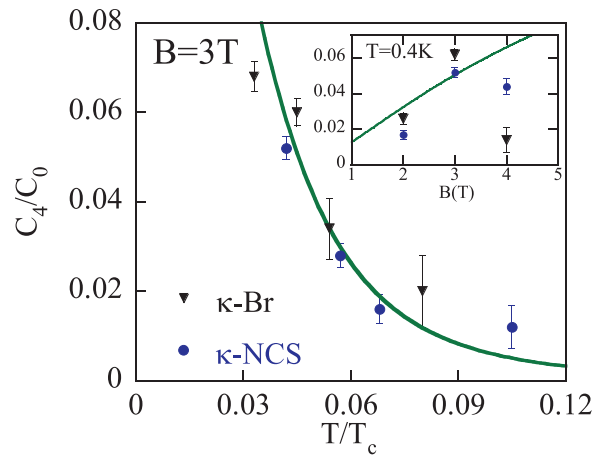}
 \caption{The normalized amplitude of the fourfold component for both $\kappa$-NCS and $\kappa$-Br as a function of
 temperature at $B=3$\,T. The inset show the field dependence at fixed $T=0.4$\,K.  The solid lines are fits to the
 nodal approximation theory (see text), the dashed lines in the inset are guides to the eye.}
 \label{FigHTdep}
\end{figure}

Data for $\kappa$-NCS are also shown in Fig.\ \ref{Figfourfold}. The magnitude of the fourfold term is comparable to
that found for $\kappa$-Br, and importantly the maxima/minima are at the same angles, so we expect the symmetry of the
gap functions to be the same. For this sample we had to introduce an additional term $C_{1}\cos(\phi+\delta_{1})$ which
is comparable in magnitude to the $C_2$ term to fully fit the background, the origin of which is unclear.  The
temperature and field dependence of the fourfold component $C_4$ for both samples is shown in Fig.\ \ref{FigHTdep}.  As
for $\kappa$-Br, $C_4/C_e$  in $\kappa$-NCS decreases strongly with increasing temperature and decreasing field.
However, for $\kappa$-NCS  $C_4/C_e$ decreases significantly less as the field is increased from B=3\,T to B=4\,T than
in $\kappa$-Br.

In Fig.\ \ref{FigHTdep} we compare our experimental data to quantitative predictions of the magnitude of this effect
for a $d$-wave superconductor.  For simplicity we use the nodal approximation of the Doppler shift theory which gives
the following expression for the field/angle dependence of the density of states
\cite{Kubert_459_1998,Vekhter_R9023_1999,BoydHVV09}
\begin{equation}
\frac{N(\omega,\bm{H},T)}{N_0}=\frac{1}{2}\left[\frac{E_1}{\Delta_0}F\left(\frac{\omega}{E_1}\right)+\frac{E_2}{\Delta_0}F\left(\frac{\omega}{E_2}\right)\right]
\end{equation}
where the field scale $E_H=0.5 a \hbar v_F\gamma_\lambda^{-\frac{1}{2}}\sqrt{\pi \mu_0 H/\Phi_0}$,
$E_1=E_H|\sin(\pi/4-\phi)|$, $E_2=E_H|\cos(\pi/4-\phi)|$ and

\[
  F(y)=
  \begin{cases}
    y\left[1+1/(2y^2)\right], & \text{if $y\geq 1$,}\\
    \left[(1+2y^2)\arcsin y + 3y\sqrt{1-y^2}\right]/y\pi, &
    \text{if $y\leq 1$.}
  \end{cases}
\]
In this approximation, which was shown to be in good agreement with more sophisticated treatments in the low
temperature/field limit \cite{Vorontsov_237001_2006}, only quasiparticles in the nodal directions are included. This
density of states was used to calculate the entropy $S$ then numerically differentiated to give the heat capacity
$C=T\partial S/\partial T$.  For the curve in the figure we set $E_H=0.13 \sqrt{\mu_0H}\,k_BT_c$.

The theory and experiment show reasonable agreement considering this contains just one fitting parameter $E_H$. In
particular, the theory explains the rapid decrease in signal with increasing temperature.  For $\kappa$-NCS, using the
in-plane Fermi velocity $v_F=1.1\times 10^5$m/s \cite{WosnitzaCWGWC91}, and penetration depth anisotropy
$\gamma_\lambda\simeq$ 100 \cite{Carrington_4172_1999}, this gives $E_H\simeq 0.17\sqrt{\mu_0H}\,k_BT_c$ (setting the
numerical constant $a=1$). This agreement with the experimental value is probably fortuitous as $a$ and the effective
value of $\gamma_\lambda$ depend on the details of the vortex lattice \cite{Vekhter_R9023_1999}.  The increase in field
between 2 and 3\,T is similar to that predicted but the theory does not explain the observed decrease in higher field.
However, more sophisticated treatments \cite{Vorontsov_237001_2006,BoydHVV09} actually predict that the oscillations in
$C$ change sign for $T\gtrsim 0.1 T_c$ or for $H\gtrsim (0.2-0.4) H_{c2}$ and so a strong decrease of $C_4$ with
increasing field is expected near these transition points.

\section{Conclusions}

In summary, the magnetic field angle dependence of the heat capacity of $\kappa$-Br and $\kappa$-NCS shows a clear
fourfold oscillation. These oscillations are shown to arise from nodes in the superconducting gap situated along the
crystal axis. These results imply a $d_{xy}$ order parameter that is consistent with the spin fluctuation mediated
pairing theory of Schmalian \cite{Schmalian_4232_1998} and others \cite{Kuroki06}. At first sight these results seem to
contradict the angular thermal conductivity $\kappa$ experiments performed on $\kappa$-NCS \cite{Izawa_027002_2002}
which showed that the small fourfold component ($\sim 0.1$\% of the total $\kappa$) was $45^\circ$ out of phase with
the one observed in this experiment (i.e., a maximum of $\kappa$ was observed with the field along the $b$ or $c$ axes
whereas a minimum is observed here).  A similar discrepancy between the $C$ and $\kappa$ measurements was found for
CeCoIn$_5$ \cite{AokiSSSOMM04,Izawa01}. Vorontsov {\it et al.} \cite{Vorontsov_237001_2006} have shown that the
oscillations in $C$ and $\kappa$ both change sign as a function of $H$ and $T$ and so the CeCoIn$_5$ results could be
explained because the measurements of $C$ and $\kappa$ were conducted in different regions of the ($H$,$T$) phase
diagram.  According to the phase diagrams in Ref.\ \onlinecite{Vorontsov_224502_2007} our $C$ measurements should be
well within the low ($T$,$H$) limit so the minima are along the nodal directions.  The $\kappa$ measurements were made
at similar $H$ and $T$ and also should be in the low ($H$,$T$) limit with minima along the nodal directions, however in
this case they are much closer to the sign switching phase boundary.  Hence the previous thermal conductivity and the
present data are not necessarily inconsistent and might be explained if there were small material dependent changes to
the global phase diagrams of Ref.\ \onlinecite{Vorontsov_224502_2007}.  Further work will be required to understand
this.

\section*{Acknowledgements}

We thank M.\ Haddow for help with x-ray diffraction and I.~Vekhter for helpful comments. This work was supported by
EPSRC (UK) and Argonne, a U.S. Department of Energy Office of Science laboratory, operated under Contract No.
DE-AC02-06CH11357.

\bibliography{organic}

\end{document}